\documentclass[a4paper]{article}

\usepackage{INTERSPEECH2021}
\usepackage{multirow}
\title{vTTS: visual-text to speech}
\name{Yoshifumi Nakano$^1$, Takaaki Saeki$^1$, Shinnosuke Takamichi$^1$,\\Katsuhito Sudoh$^2$, Hiroshi Saruwatari$^1$}
\address{
  $^1$The University of Tokyo, Japan. $^2$Nara Institute of Science and Technology, Japan.}
\email{nakano-yoshifumi230@g.ecc.u-tokyo.ac.jp, shinnosuke\_takamichi@ipc.i.u-tokyo.ac.jp}

\begin{document}

\maketitle

\begin{abstract}
    This paper proposes visual-text to speech (vTTS), a method for synthesizing speech from visual text (i.e., text as an image). Conventional TTS converts phonemes or characters into discrete symbols and synthesizes a speech waveform from them, thus losing the visual features that the characters essentially have. Therefore, our method synthesizes speech not from discrete symbols but from visual text. The proposed vTTS extracts visual features with a convolutional neural network and then generates acoustic features with a non-autoregressive model inspired by FastSpeech2. Experimental results show that 1) vTTS is capable of generating speech with naturalness comparable to or better than a conventional TTS, 2) it can transfer emphasis and emotion attributes in visual text to speech without additional labels and architectures, and 3) it can synthesize more natural and intelligible speech from unseen and rare characters than conventional TTS.
\end{abstract}

\noindent\textbf{Index Terms}: speech synthesis, visual text, visual-text to speech, visual feature

\section{Introduction}
\label{sec:introduction}
\begin{figure}[t]
        \centering
        \includegraphics[width=7cm]{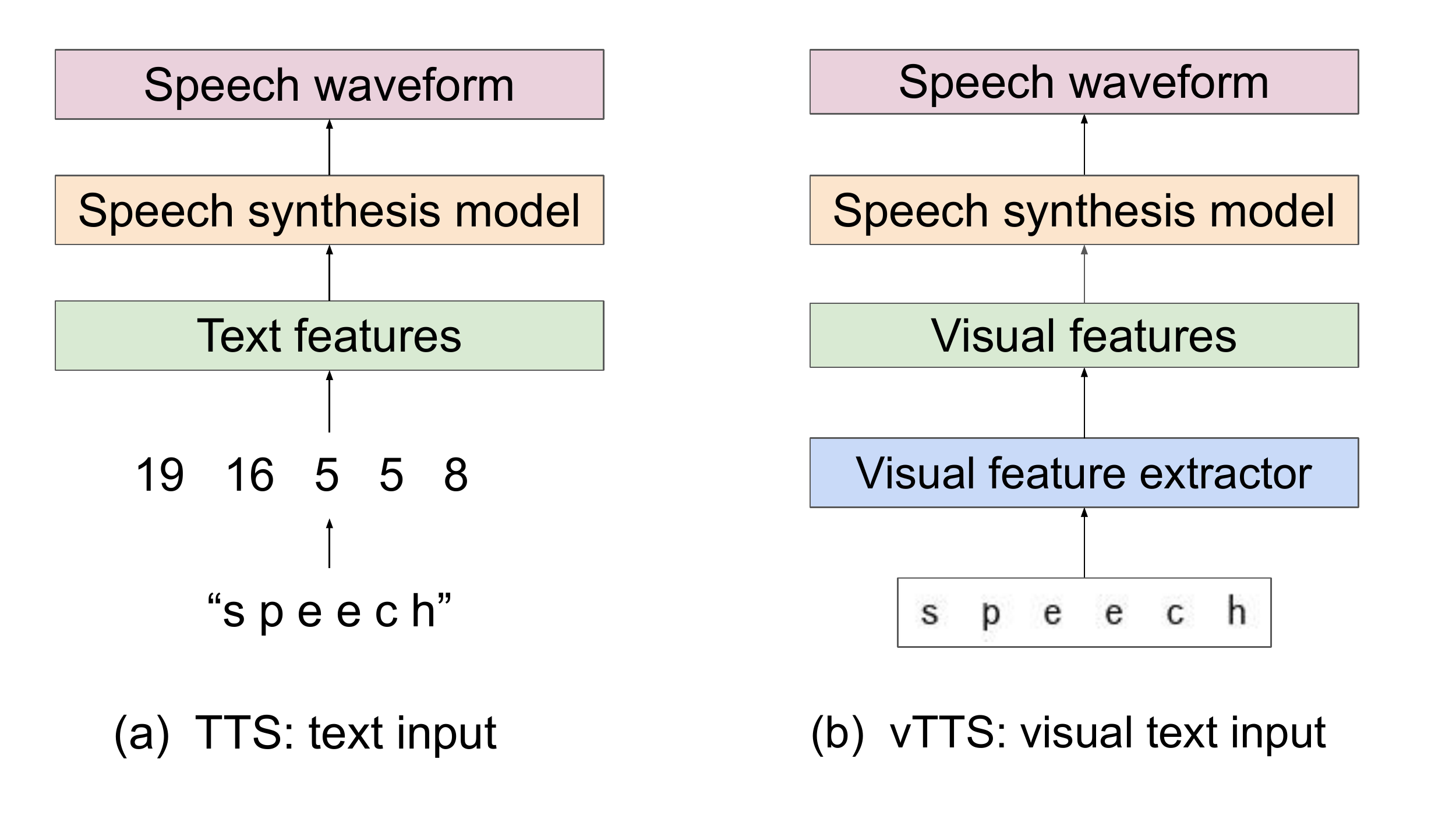}
        \vspace{-4mm}
        \caption{Comparison of speech synthesis methods. Whereas the basic TTS uses discrete text features, our vTTS uses visual features.} 
        \vspace{-2mm}
        \label{fig:figure1}
    \end{figure}

    Text to speech (TTS) is a method for synthesizing speech from arbitrary text. Along with the development of deep neural networks (DNNs), neural TTS has succeeded in synthesizing speech with the same naturalness as human speech~\cite{WaveNet,Tacotron,Tacotron2,fastSpeech}. Generally, TTS first converts each character (or each phoneme based on linguistic knowledge) into a discrete symbol and synthesizes a speech waveform from it~(Figure~\ref{fig:figure1}(a)).
    
    However, when we read a sentence out loud, we do not view each character as a discrete symbol but rather use the \textit{visual information} of the character. For example, when we read a phonogram (a character representing a speech sound), the character or the combination of sub-characters (components of one character) determines the reading. In addition, we can recognize attributes in visual text and reflect them in speech. For example, when reading a textbook, we can recognize an underlined (or bold, italic) word as important and read the word emphatically~\cite{font_emp}. Another example is typeface, which sometimes evokes a certain emotion~\cite{English_typeface,japanese_font}. Advertisements and comics utilize this to convey desired emotions to readers~\cite{application_to_music}. From these facts, it is more appropriate to use the visual information of text to synthesize speech.
    
    In this paper, we propose visual-text to speech (vTTS), which generates a speech waveform from visual text (Figure~\ref{fig:figure1}(b)). Our vTTS consists of a visual feature extractor and FastSpeech2-inspired model. The visual feature extractor extracts visual features from visual text using a convolutional neural network (CNN). The FastSpeech2-inspired model predicts acoustic features from the extracted visual features. 
    
    Our vTTS has three advantages over conventional TTS. First, vTTS can reflect emphasis and emotion attributes in speech by recognizing the difference of the typeface of visual text itself whereas the basic TTS requires additional labels and architectures~\cite{emotion-tts,emotion-tts2}. Second, our vTTS does not require a vocabulary (lookup table) of characters, unlike the basic TTS. A pre-determined fixed-size vocabulary cannot be used to embed out-of-vocabulary (OOV) characters (typically, ``unknown'' symbol is used). Our vTTS avoids this and obtains visual features from OOV characters. Last, for some languages, our model can efficiently predict the reading from character compositionality (the reading or meaning is determined by the structure of the character). Even if rare and OOV characters emerge, vTTS can synthesize more natural and intelligible speech utilizing character compositionality than the basic TTS. 
    
    We compared our vTTS with the basic TTS in three phonetic languages: Japanese (Hiragana), Korean (Hangul), and English (Roman Alphabet) and then revealed the following properties.
    \vspace{-1mm}
    \begin{itemize} \leftskip -5mm \itemsep -1mm
        \item Our vTTS is comparable to or better than the basic TTS in terms of naturalness of synthetic speech. 
        \item It can transfer emphasis and emotion attributes in visual text to speech without additional labels and architectures.
        \item It can synthesize more natural and intelligible speech from OOV and rare phonograms than the basic TTS.
    \end{itemize}
    \vspace{-1mm}
    
    Our implementation\footnote{https://github.com/Yoshifumi-Nakano/visual-text-to-speech} and corpus\footnote{https://sites.google.com/site/shinnosuketakamichi/research-topics/jecs\_corpus} are open-sourced on the project pages for reproducibility and future work.

\vspace{-2mm}
\section{Related Work}
    \subsection{Text to speech (TTS)}
    \vspace{-2mm}
    \label{subsec:TTS}
        As the basic TTS model, we use FastSpeech2~\cite{fastspeech2}, a well-known non-autoregressive model. FastSpeech2 consists of three components: an encoder, variance adapter, and decoder. The input of the original FastSpeech2 is a phoneme sequence, but we use a character sequence instead for fair comparison with the proposed method. Section \ref{subsec:setup} notes the results of preliminary experiments on this change.
        
        Some studies use information other than text such as reference voices~\cite{reference-voice}, face images~\cite{face-images}, and articulation~\cite{face-articulation}. These methods have an additional encoder for such information. However, similar architecture is not suitable for vTTS. Since visual text is not discrete as described in Section \ref{sec:introduction}, text and visual information should not be treated separately. Therefore, we propose a model that jointly treats text and visual information in Section \ref{sec:method}.
        
        Although a character is a input unit of speech synthesis that does not rely on language knowledge, some studies use other units, e.g., phrase~\cite{sp-phrase}, word~\cite{sp-word}, subword~\cite{sp-subwoed}, and byte~\cite{sp-byte} in descending order of length. In this paper, we use a pixel, which is a smaller input unit than ever before. We can observe a similar trend in natural language processing as described below.
        
    \vspace{-2mm}
    \subsection{Visual text for natural language processing (NLP)}
    \vspace{-2mm}
        In NLP, the input unit of language models and machine translation models has been examined. Starting with phrases~\cite{nlp-pharase} and words~\cite{nlp-word} and moving through subwords~\cite{nlp-subword}, characters~\cite{nlp-character}, and sub-characters~\cite{nlp-subcharacter}, some research proposes the use of pixels (i.e., visual text).
        
        There is one study that extracted character embeddings from visual text~\cite{Compositionality}. This study utilized logogram compositionality, allowing for better processing of rare words. Similarly, there are studies that have used character compositionality in a word segmentation task~\cite{glygh-chinese} or a sentiment classification task~\cite{super-character}. Other than that, the use of sliced visual text, which is analogous to subword text symbols, leads to there being more resistance to various types of noise~\cite{open-vocab}. These studies mainly used a CNN to extract the visual information from visual text. 
        
        These studies are based on the compositionality in logograms and acquire meaning-related features. In contrast, our study aims to acquire sound-related ones contained in characters, especially in phonograms.

\vspace{-2mm}
\section{Method}
\label{sec:method}
    \begin{figure}[t]
        \centering
        \includegraphics[width=7cm]{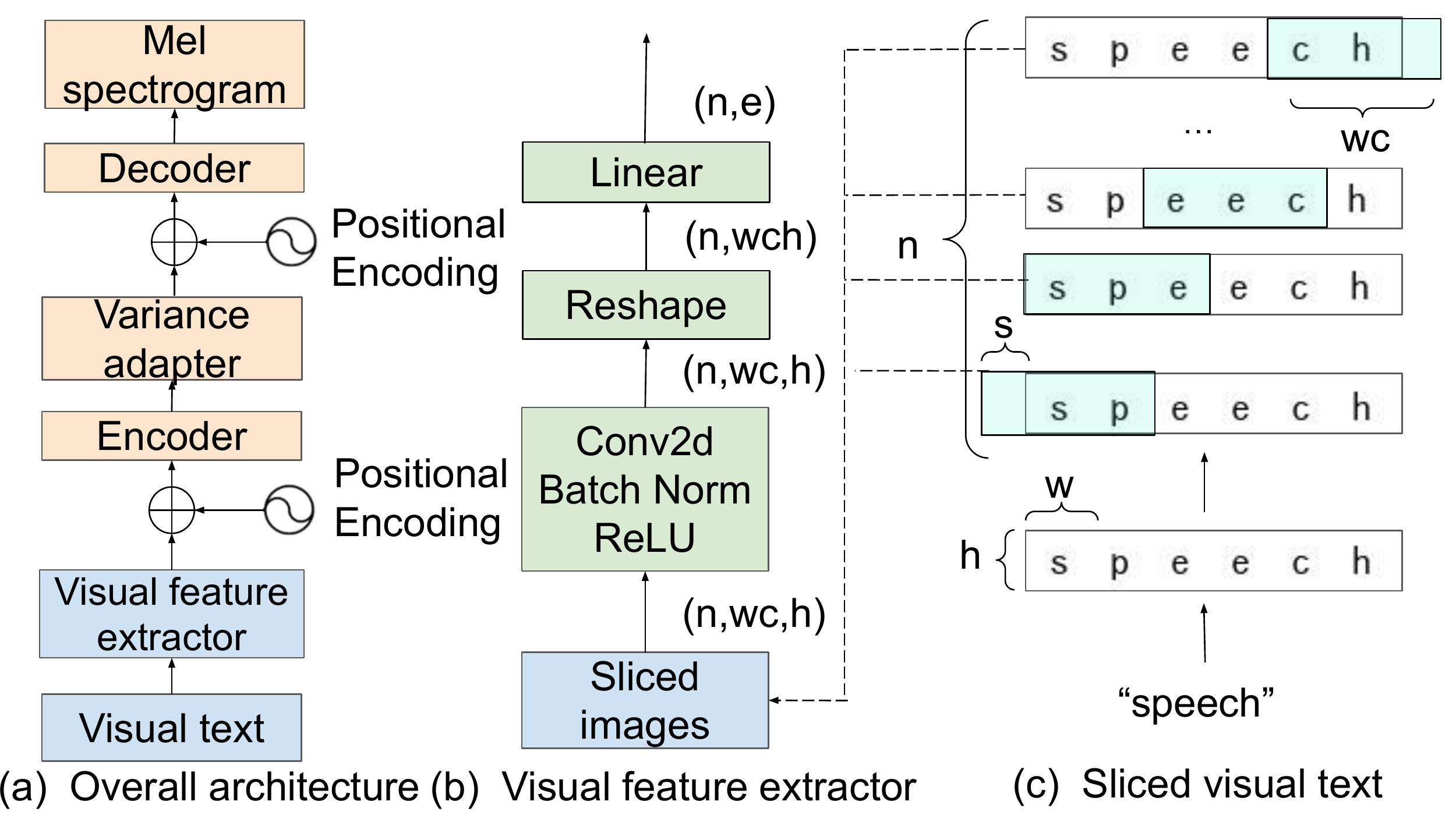} 
        \vspace{-2mm}
        \caption{Overall architecture of the proposed method.}
        \vspace{-2mm}
        \label{fig:figure2}
    \end{figure}
    
    \begin{figure}[t]
        \centering
        \includegraphics[width=7cm]{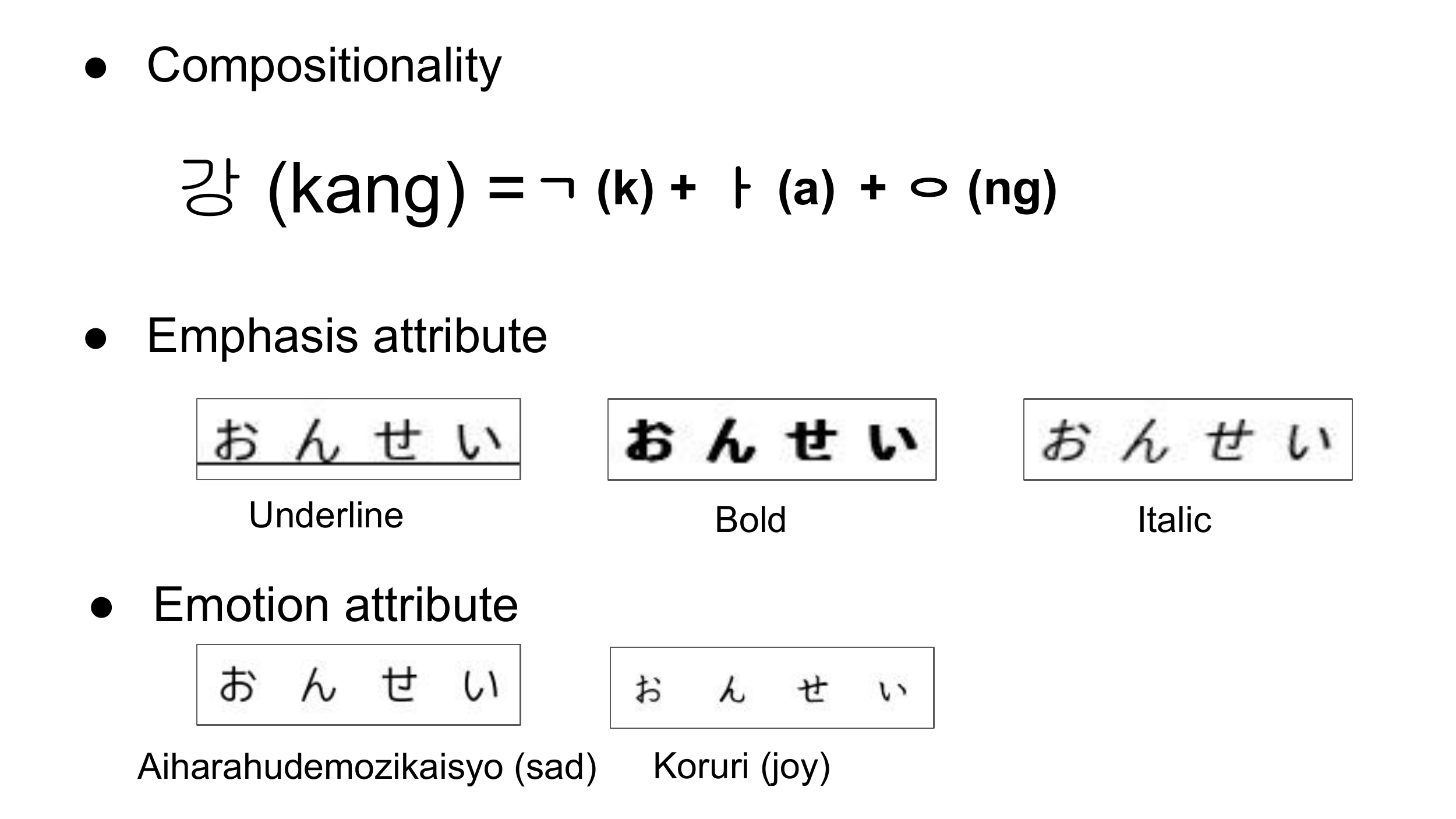} 
        \vspace{-2mm}
        \caption{Example of visual text. Structure and combination of sub-characters determine the overall reading (top). Emphatic (middle) and typeface-decorated (bottom) text evoke emphasis and emotion, respectively.}
        \vspace{-2mm}
        \label{fig:figure3}
    \end{figure}
    \vspace{-2mm}
    \subsection{Architecture comparison between vTTS and TTS}
    \vspace{-2mm}
         As for the architecture of the proposed vTTS, we replace the character embedding layer (i.e., text feature extractor) of the original FastSpeech2 with a visual feature extractor (Figure~\ref{fig:figure2}(a)). This replacement allows for extracting visual features from visual text. The subsequent encoder, variance adapter, and decoder are the same as the original FastSpeech2. To investigate the ideal performance of vTTS, in this paper, we use artificially generated visual text rather than that in in-the-wild images (e.g., advertisements and comics).
         
    \vspace{-2mm}
    \subsection{Generating visual text}
    \vspace{-2mm}
    \label{subsec:GeneratingVisualText}
        The first step is to transform text to grayscale sliced images. The overall process is illustrated in Figure~\ref{fig:figure2}(c). Since the original FastSpeech2 is a non-autoregressive TTS utilizing a duration-based upsampler, we must take the temporal alignment between visual text and a speech feature sequence. Therefore, we use visual text with monospace fonts in this work. Each character is of a specified width $w$, height $h$, and font size $fs$. Therefore, characters of length $n$ are transformed into an image whose width is $nw$ and height is $h$. We extract sliced images from it using sliding windows similar to a previous study~\cite{open-vocab}. The window is of a specified width $wc$ and height $h$ extracted at intervals $s(=w)$. Here, $c$ represents the number of characters in one sliced image, and adjusting $c$ leads to a change in the preceding and following characters to be considered when obtaining visual features from each piece of visual text. For example, if $c=3$, convolution is performed including one character before and after. By considering the number of preceding and following characters, reading and prosody can be acquired that depend on the neighboring characters. Blank images are padded to the left and right edges to create sliced images for the number of characters. Finally, we generate $n$ sliced images of height $h$ and width $wc$ from text.
    
    \vspace{-2mm}
    \subsection{Visual features}
    \vspace{-2mm}
        We extract visual features from the sliced images generated in Section~\ref{subsec:GeneratingVisualText}. The basic TTS uses the output from a text feature extractor. In comparison, vTTS uses the output of a visual feature extractor instead. The architecture of the extractor is shown in Figure~\ref{fig:figure2}(b) and is inspired by a machine translation study using visual text~\cite{open-vocab}. The extractor mainly comprises $2$D convolution, $2$D batch normalization~\cite{pmlr-v37-ioffe15}, and the activation function ReLU~\cite{Nair2010RectifiedLU}. Finally, we apply a linear layer to the reshaped output to obtain a visual feature for each character. The visual features are fed to the FastSpeech2-inspired encoder.
        
        The visual feature extractor utilizes the following information. Figure~\ref{fig:figure3} shows examples of visual text with the following information.
        \begin{itemize} \leftskip -5.5mm \itemsep 0mm
            \item \textbf{Compositionality:}~In phonetic languages (e.g., Korean), the character or combination of sub-characters determines the overall reading. The visual feature extractor acquires the correspondence between a character (or sub-character) and the reading. Therefore, even if OOV and rare characters emerge, we expect our vTTS to be able to accurately predict the readings using the visual features of the characters or sub-characters.
            \item \textbf{Emphasis attribute:}~The visual feature extractor distinguishes emphasized typefaces (underline, bold, italic) from normal typefaces. This enables vTTS to synthesize word-emphasized speech only from emphasized typefaces.
            \item \textbf{Emotional attribute:}~The visual feature extractor acquires the correspondence between typeface and emotion, and it reflects the emotion in speech in accordance with the typeface. 
        \end{itemize}
    
\section{Experiments and Results}
    \subsection{Experimental Setup}
    \label{subsec:setup}
        \textbf{Language.}~We evaluated the performance of the proposed method on three phonetic languages: Japanese (Hiragana), Korean (Hangul), and English (Roman Alphabet). For Japanese, we converted Katakana and Kanji into $80$ Hiragana-characters in advance. In Hiragana, one character corresponds to the specific sound.  For Korean, we used $1226$ Hangul-characters. In Hungul, the combination of jamo determines the sound. For English, we applied lowercasing to facilitate training. 
        
        \noindent\textbf{Dataset.}~We used the JSUT~\cite{JSUT} (Japanese), KSS~\cite{KSS} (Korean), and LJSpeech~\cite{ljspeech} (English) corpora for naturalness comparison of vTTS and TTS as described in Section~\ref{subsubsec:audio-quality}. The training (validation) data sizes for these languages were $8.3$ ($0.61$), $9.0$ ($0.38$), and $19$ ($0.89$) hours, respectively. The test data sizes were $100$ samples for all languages. We used the JECS corpus (see our project page), which contains noun-emphasized Japanese speech, for the emphasis experiments described in Section~\ref{subsubsec:Emp}. The training, validation, and test data sizes were $0.412$ hours, $0.03$ hours, and $50$ samples, respectively. We used the manga2voice corpus~\cite{manga2voice}, which contains emotional speech, for the emotion experiment described in Section~\ref{subsubsec:Emo}. The training, validation, and test data sizes were $0.062$ hours, $0.003$ hours, and $50$ samples, respectively. Since the training data sizes of the JECS and manga2voice corpora were small, we fine-tuned a vTTS model pretrained using the JSUT corpus. We used the KSS corpus in Section~\ref{subsubsec:openV}. The data sizes were almost the same but slightly different from those in Section~\ref{subsubsec:audio-quality}. See Section~\ref{subsubsec:openV} for details. All audio was downsampled to $22.05$ kHz. The alignment between visual text and acoustic features was obtained through forced alignment of visual text and a character sequence, and a character sequence and a acoustic feature sequence. 
        
        \noindent\textbf{Visual text.}~We used pygame\footnote{https://www.pygame.org/news} to generate visual text from text. For Japanese and English, we used the IPA typeface; for Korean, the Gowun Batang typeface. Visual text of $w=30$ and $h=30$ was used for all languages. $fs$ was set to $15$ for Japanese and Korean and $20$ for English. This is the result of setting the text to be the appropriate size for an image. 
        
        \noindent\textbf{Model configuration.}~The model size and hyperparameters of the original FastSpeech2 and FastSpeech2-inspired architecture in vTTS followed the open-source implementation~\cite{multi-FastSpeech2}\footnote{https://github.com/Wataru-Nakata/FastSpeech2-JSUT}. In the visual feature extractor, the $2$D convolution used only one channel, and there was a padding of $1$, kernel size of $3$, and stride of $1$~\cite{open-vocab}. The dimension of the output of the visual feature extractor was set to $256$, the same as the encoder hidden dimension of FastSpeech2. We generated speech waveforms from mel spectrograms using the pretrained HiFi-GAN~\cite{HiFi-GAN}\footnote{https://github.com/jik876/hifi-gan}. 
        
        \noindent\textbf{Preliminary experiment.}~As described in Section \ref{subsec:TTS}, we used characters instead of phonemes as input unit for FastSpeech2. We conducted a subjective evaluation of this substitution under the same conditions as in Section~\ref{subsubsec:audio-quality}. The results showed that there was no significant difference in naturalness in Japanese and English. On the other hand, the use of phonemes was better than that of characters and visual text in Korean. This comparison is not the main scope of this paper, but deep investigation will be our future work.
        
    \vspace{-2mm}
    \subsection{Results}
        \subsubsection{Naturalness comparison between vTTS and TTS}
        \vspace{-2mm}
        
        \label{subsubsec:audio-quality}
            \begin{table}[t]
                \centering
                \setlength{\tabcolsep}{1.5mm}
                \caption{Results of MOS and $95\%$ confidence interval for naturalness. \textbf{Bold} means significantly better score than TTS. For all languages, vTTS was comparable to or better than TTS in naturalness.}
                \label{tab:audio-quality}
                \vspace{-2mm}
                \footnotesize
                \begin{tabular}{c|cccc}
                \multicolumn{1}{c|}{\multirow{2}{*}{Lang.}} & \multicolumn{1}{c|}{\multirow{2}{*}{TTS}} & \multicolumn{3}{c}{vTTS}                                   \\ \cline{3-5} 
                \multicolumn{1}{c|}{}                       & \multicolumn{1}{c|}{}                     & \multicolumn{1}{c|}{$c=1$} & \multicolumn{1}{c|}{$c=3$} & $c=5$ \\ \hline
                ja                                          & $3.45 \pm 0.09$                           & $3.41 \pm 0.09$            & $3.46 \pm 0.09$            & $3.49 \pm 0.10$ \\
                ko                                          & $3.04 \pm 0.16$                           & $\mathbf{3.55 \pm 0.15}$   & $3.18 \pm 0.15$            & $3.01 \pm 0.15$ \\
                en                                          & $3.72 \pm 0.10$                           & $3.69 \pm 0.10$            & $3.70 \pm 0.11$            & $3.71 \pm 0.10$
                \end{tabular}
                \vspace{-2mm}
            \end{table}
            
            To compare our vTTS with conventional TTS, we conducted a mean opinion score (MOS) evaluation on the naturalness of the synthetic speech. We used three settings for vTTS, $c=1, 3, 5$, to investigate the impact of the window size. Native speakers of each language participated in the evaluation. $150$ and $200$ speakers recruited using our crowdsourcing system listened to $20$ speech utterances in Japanese and English, respectively. In contrast, $10$ speakers recruited by snow-ball sampling\footnote{We could not collect enough number of native listeners in our crowdsourcing system} listened to $160$ speech utterances in Korean. The text content was kept consistent among different methods so that all listeners examined only naturalness without other interfering factors.

            The results are shown in Table~\ref{tab:audio-quality}. For all languages, the naturalness of vTTS was comparable to or better than that of TTS. Comparing the highest score of vTTS ($c=5$ for Japanese and English, $c=1$ for Korean) with TTS, although there was no significant difference for Japanese and English, there was a significant difference for Korean ($p<0.05$). This means that the extraction of visual features was effective for Hangul, whose readings are determined by the combination of sub-characters. 
            
            It can also be seen that the appropriate window size differed depending on the language. For English, there was no significant difference between $c=1,3,5$. However, for Japanese, the naturalness with a sliding window of $c=5$ was slightly better than that with $c=1$ ($p<0.10$), while for Korean, $c=1$ was significantly better than $c=5$ ($p<0.001$). One possible reason is that the number of phonemes expressed by one character differs depending on the language. The following experiments used the best $c$ value for each language.

        \vspace{-2mm}
        \subsubsection{Emphasis attribute}
        \label{subsubsec:Emp}
        \vspace{-2mm}
           We evaluated whether our vTTS can transfer the emphasis attribute in visual text to speech. We trained the vTTS model using pairs of word-emphasized speech and visual text. We used three types of word emphasis methods (underline, bold, and italic) as shown in Figure~\ref{fig:figure3}(b). We trained the vTTS model using each type, i.e., three vTTS models were trained. The trained models try to synthesize speech with emphasis placed on specified words. We prepared synthetic speech of not only emphasis groups (``Emphasis (*)'') but also control groups (``Control (*)''). Their difference was in the input visual text: word-emphasized visual text in the emphasis groups and non-emphasized text in the control groups. 

            For the listening test, listeners listened to speech randomly selected from seven kinds (three emphasis groups, three control groups, and ground-truth natural speech) of speech utterances. The listeners answered with which word was emphasized, from word (specifically, noun) candidates. $150$ native speakers listened to $20$ speech utterances. The average number of candidates per sentence was $3.6$; the chance rate was $0.299$. 
            
            The results are shown in Table~\ref{tab:emp}. For all emphasis methods, the accuracy of the emphasis groups was significantly better than that of the control groups (${p<0.01}$). This shows that our vTTS properly recognized the emphasis attribute in visual text. In addition, the accuracy of the emphasis groups was close to that of the ground truth. There was no significant difference between two emphasis methods (underline, bold) and the ground truth. These results show that these two emphasis methods can accurately transfer the emphasis attribute in visual text to speech, without additional emphasis labels and architectures. One possible reason for the lower accuracy of the italic input compared with the two other emphasis methods is that the shape of italic is not as similar to that of normal typeface because of geometric transformation.

            \begin{table}[t]
                \centering
                \caption{Accuracy of perceived emphasized words. Our method can properly reflect emphasis attribute in visual text in speech.}
                \vspace{-2mm}
                \footnotesize
                \label{tab:emp}
                \begin{tabular}{c|c}
                Speech & Accuracy\\ 
                \hline\hline
                Ground truth           & $0.960$   \\ 
                \hline
                Emphasis (underline) & $0.933$   \\ 
                \hline
                Emphasis (bold)      & $0.898$  \\ 
                \hline
                Emphasis (italic)      & $0.877$  \\ 
                \hline
                Control (italic)     & $0.505$  \\ 
                \hline
                Control (bold)          & $0.400$   \\ 
                \hline
                Control (underline)     & $0.381$  \\ 
                \end{tabular}
                \vspace{-2mm}
            \end{table} 
        
        \vspace{-2mm}
        \subsubsection{Emotion attribute}
        \label{subsubsec:Emo}
        \vspace{-2mm}
            \begin{table}
                \centering
                \caption{Confusion matrix between reference (true\_*) and perceived (pred\_*) emotion. Our method can properly reflect emotion attributes in visual text in synthetic speech.}
                \vspace{-2mm}
                \footnotesize
                \label{tab:emotion}
                \begin{tabular}{c|cc}
                             & pred\_happy & pred\_sad  \\ 
                \hline
                true\_happy & $0.795$        & $0.205$       \\
                true\_sad    & $0.114$        & $0.886$     
                \end{tabular}
                \vspace{-2mm}
            \end{table}
            
            We evaluated whether our method can transfer emotions evoked by certain typefaces to speech. We trained the vTTS model using pairs of emotional speech and typeface-decorated visual text. Before the training, we conducted preliminary experiments to find typefaces that evoke emotions. As a result, we decided to use the Koruri typeface\footnote{https://github.com/Koruri/Koruri/releases/tag/20210720} and Aiharahudemozikaisyo typeface\footnote{https://faraway.work/font.html}, which evoke joy and sadness, respectively. These are typefaces that more than $70$ percent of evaluators thought of as corresponding emotion. 
            
            We trained one vTTS model using the above two typefaces. In the listening test, listeners listened to the synthetic speech and answered the perceived emotion. To avoid the listener from perceiving emotion from the linguistic content, the content was common between emotions, and it was selected from the ``neutral'' subset of the manga2voice corpus. $120$ listeners participated, and each listener listened to $20$ speech utterances.
            
            The results are shown in Table~\ref{tab:emotion}. When listening to speech synthesized from the Koruri typeface, $79$\% of the listeners answered that it sounded joyful, and $89$\% of the listeners answered that it sounded sad when listening to speech synthesized from the Aiharahudemozikaisyo typeface. These scores are comparable to TTS with emotion labels~\cite{emotion-tts}. These results show that vTTS can accurately transfer the emotion attribute in visual text to speech, without additional emotion labels and architectures.

        \vspace{-2mm} 
        \subsubsection{Speech synthesis from rare and OOV characters}
        \vspace{-2mm}
            \label{subsubsec:openV}
            \begin{table}[t]
    \centering
    \setlength{\tabcolsep}{1.3mm}
    \caption{Results of MOS and $95\%$ confidence interval on naturalness. $\Delta$ denotes MOS decrease from ``in-vocab.'' Our method can synthesize more natural speech than TTS.}
    \vspace{-3mm}
    \footnotesize
    \label{tab:openv-natural}
    \begin{tabular}{c|ccc}
            & in-vocab              & rare ($\Delta$)               & OOV ($\Delta$) \\ \hline
    TTS     & $3.29 \pm 0.16$       & $2.32 \pm 0.16$ ($-0.97$)     & $2.31 \pm 0.20$ ($-0.98$)   \\
    vTTS    & $3.58 \pm 0.13$       & $3.12 \pm 0.16$ ($-0.46$)     & $2.95 \pm 0.21$ ($-0.63$)   \\
    \end{tabular}
\end{table}
            
            \begin{table}[t]
    \centering
    \caption{Results on CER. $\Delta$ denotes CER increase from ``in-vocab.'' Our method can synthesize more intelligible speech than TTS.}
    \vspace{-2mm}
    \footnotesize
    \label{tab:openv-cer}
    \begin{tabular}{c|ccc}
            & in-vocab      & rare ($\Delta$)               & OOV ($\Delta$) \\ \hline
    TTS     & $0.120$       & $0.194$ ($+0.074$)            & $0.255$ ($+0.135$)   \\
    vTTS    & $0.080$       & $0.114$ ($+0.034$)            & $0.163$ ($+0.083$)   \\
    \end{tabular}
\end{table}    
                
            Finally, we investigated the ability of vTTS to synthesize speech from OOV and rare characters. A good example for this purpose is Korean because the character compositionality determines the reading. We designed three test sets: in-vocab, rare, and OOV. The ``in-vocab'' set consisted of only characters appearing more than three times in the training data. Sentences in the ``rare'' set included characters appearing less than three times in the training data, and those in the ``OOV'' set included OOV characters. Each test set consisted of $50$ sentences. For comparison, we trained the TTS model under the same conditions as for the vTTS. The TTS model had an ``unknown'' symbol, and an OOV character was encoded in the symbol. We conducted a MOS test on naturalness and a dictation test. The listener listened to speech utterances randomly selected from all test sets and then answered the speech naturalness in the MOS test and dictated the text in the dictation test. The character error rate (CER) was used in the dictation test. $10$ listeners participated in each test. Each listener listened to $160$ and $64$ speech utterances in the MOS test and dictation test, respectively. 

            The MOS results on naturalness are shown in Table~\ref{tab:openv-natural}. There was a significant difference between vTTS and TTS in the ``in-vocab,'' ``rare,'' and ``OOV'' sets ($p<0.001$). For both TTS and vTTS, the scores decreased from ``in-vocab'' to ``rare'' and ``OOV,'' but the changes in vTTS were smaller than those in TTS. The dictation results for intelligibility are shown in Table~\ref{tab:openv-cer}. The overall tendency was the same as the MOS test. There was a significant difference between TTS and vTTS in the ``in-vocab,'' ``rare,'' and ``OOV'' sets ($p<0.001$). Also, the CER increase for vTTS was also smaller than that for TTS. 
            
            From these results, we can say that our vTTS can capture character compositionality for speech synthesis and synthesize intelligible speech even from OOV and rare characters.

\vspace{-2mm}        
\section{Conclusions}
\vspace{-2mm}
    In this paper, we proposed vTTS, which generates speech waveforms from visual text. Experimental results showed that 1) the basic performance of vTTS is comparable to or better than that of TTS, 2) it can reflect several attributes in visual text in speech, and 3) it can generate more natural speech from OOV and rare characters. In future work, we plan to apply our method to a variety of media such as manga or posters. We hope that this research will serve as a first benchmark in the development of speech synthesis from visual text.

\textbf{Acknowledgements:} This work was supported by JST, Moonshot R\&D Grant Number JPMJPS2011 (for multi-language experiments), and JSPS KAKENHI 21H05054, 19H01116, 21H04900 (for basic technique).

\bibliographystyle{IEEEtran}



\end{document}